\documentclass[conference]{IEEEtran}

\IEEEoverridecommandlockouts

\usepackage{amsmath,amssymb,amsfonts}
\usepackage{algorithmic}
\usepackage{graphicx}
\usepackage{textcomp}
\usepackage{xcolor}

\def\BibTeX{{\rm B\kern-.05em{\sc i\kern-.025em b}\kern-.08em
    T\kern-.1667em\lower.7ex\hbox{E}\kern-.125emX}}

\newcommand*\diff{\mathop{}\!\mathrm{d}}

\usepackage{url} 

\usepackage{bm}
\usepackage[normalem]{ulem} 
\usepackage{supertabular}
\usepackage{amsthm}
\theoremstyle{definition}

\usepackage[linesnumbered,ruled,vlined]{algorithm2e}
\usepackage{pbox}
\usepackage{float} 
\usepackage{subcaption} 

\usepackage{cite} 

\begin{document}

\title{Optimizing RRH Placement Under a \\ Noise-Limited Point-to-Point Wireless Backhaul}
\author{Hussein~A.~Ammar\IEEEauthorrefmark{1}, 
	Raviraj~Adve\IEEEauthorrefmark{1}, Shahram~Shahbazpanahi\IEEEauthorrefmark{2}, and Gary~Boudreau\IEEEauthorrefmark{3}\\ 
	\IEEEauthorrefmark{1}University of Toronto, Dep. of Elec. and Comp. Eng., Toronto, Canada\\
	\IEEEauthorrefmark{2}University of Ontario Institute of Technology, Dep. of Elec. and Comp. Eng., Oshawa, Canada\\
	\IEEEauthorrefmark{3}Ericsson Canada, Ottawa, Canada\\
	Email: \{ammarhus, rsadve\}@ece.utoronto.ca
}

\maketitle

\begin{abstract}
In this paper, we study the deployment decisions and location optimization for the remote radio heads (RRHs) in coordinated distributed networks in the presence of a wireless backhaul. We implement a scheme where the RRHs use zero-forcing beamforming (ZF-BF) for the access channel to jointly serve multiple users, while on the backhaul the RRHs are connected to their central units (CUs) through point-to-point wireless links. We investigate the effect of this scheme on the deployment of the RRHs and on the resulting achievable spectral efficiency over the access channel (under a backhaul outage constraint). Our results show that even for noise-limited backhaul links, a large bandwidth must be allocated to the backhaul to allow freely distributing the RRHs in the network. Additionally, our results show that distributing the available antennas on more RRHs is favored as compared to a more co-located antenna system. This motivates further works to study the efficiency of wireless backhaul schemes and their effect on the performance of coordinated distributed networks with joint transmission. 
\end{abstract}

\section{Introduction}
Cooperation between the network transmitters is imperative to control interference. In a distributed network, this can be achieved by deploying two key modules: a central unit (CU) that executes protocols to schedule the transmission and the reception of the signals, and remote radio heads (RRHs) deployed throughout the network's coverage area to cooperate and jointly serve users~\cite{3GPP:TR21.915}. However, a key component for such a scheme is the backhaul links which connect RRHs to their CUs (also called fronthaul of the CU). Notably, this scheme is a distributed implementation of the 5G next-generation NodeB specified in the New Radio specification~\cite{3GPP:TR21.915}.

\par The backhaul links carry the heaviest communication load since they carry the data for all users served by the RRHs and hence are the bottleneck to throughput. The works in~\cite{7120102,7981320} modeled the reliability of dedicated backhaul links as a Bernoulli distribution to study their effect on the system outage. The aim from these studies is to derive optimal strategies for the assignment of the backhaul links. Important issues such as power allocation~\cite{8678745}, cooperation strategies~\cite{6571307}, beamforming design~\cite{7952837}, resource allocation~\cite{7581201} and formation of serving clusters~\cite{6920005} have been investigated under the theme of a limited-capacity backhaul. Additionally, the work in~\cite{8422865} analyzed backhaul signal compression as a mean to minimize the impact of the limited capacity of the~backhaul. 

As an alternative, several investigations have considered the problem of transmitter placement. The work in~\cite{8529184} studied RRHs placement without accounting for the backhaul. Furthermore, the study in~\cite{6702841} optimized the placement of relays in in-band or out-of-band non-cooperative cellular networks by modeling users traffic using queueing theory; their results show that the latter provides better performance. The work in~\cite{7161398} also optimized the deployment of relays, while the work in~\cite{8891140} studied inter-site distance for RRHs deployment on a highway road scenario, where the outage probability was derived. Nonetheless, these works about transmitters deployment did not consider the effect of the backhaul on the deployment decisions in a coordinated distributed network.

The backhaul capacity strongly affects the network performance and the flexibility of the deployment of the RRHs. Motivated by this fact, we herein investigate the effect of using point-to-point wireless backhaul on the placement of the RRHs, where the aim is to enhance the access channel spectral efficiency while meeting a backhaul outage constraint. Our work adds to the literature on both the limited-capacity backhaul and transmitters placement for distributed networks and provides insights into RRH deployment decisions. The rest of the paper is organized as follows: in Section~\ref{sec:systemModel}, we present our system model. While in Section~\ref{sec:problemFormulation}, we formulate the RRHs location optimization problem. In Sections~\ref{sec:RRHsPlacement} and~\ref{sec:results}, we present our proposed solution and results, respectively. Finally, we present our conclusions in Section~\ref{sec:Conclusion}.
\section{System Model}\label{sec:systemModel}
\subsection{Network Model}
We consider a cooperative distributed network comprising $Q$ cells. Assuming joint transmission with disjoint clustering, the RRHs found in each cell $q$ jointly serve the users inside their cell boundary $\mathcal{B}_q$~\cite{PDPUsercentricVsDisjoint8969384}. Each cell employs $N$ RRHs to serve $K$ users on the same time-frequency resource block using zero-forcing beamforming (ZF-BF). Each RRH uses $M$ antennas to serve the users, where $NM > K$. In each cell, a single CU, located at the cell center, controls the transmissions of the RRHs through point-to-point backhaul links. The users in the cell may be concentrated in (but not restricted) hotspots with \emph{a priori} known traffic pattern. The backhaul and access links are spectrally orthogonal. To focus on network design, we assume perfect channel state information (CSI) is available. Our aim is to deploy the RRHs in the network so that we obtain the highest spectral efficiency over the access channel while controlling backhaul outage.
\subsection{Access Channel Transmission Scheme}
\par The signal received at a typical user $k$ located in cell $q$ is
\vspace{-0.5em}
\begin{align} \label{eq:signalModel_accessChannel}
&r_{qk} = \displaystyle \underbrace{\sum_{n = 1}^{N}{\bf h}_{qn,qk}^H {\bf w}_{qnk} s_{qk}}_{\text{useful signal}}
+\underbrace{\sum_{\substack{n=\{1,\dots,N\},\\k'=\{1,\dots,K\}, k'\ne k}}{\bf h}_{qn,qk}^H {\bf w}_{qnk'} s_{qk'}}_{\substack{\text{intra-cluster}\text{ interference}}}\nonumber\\
&\quad
+\underbrace{\sum_{\substack{q'=\{1,\dots, Q\},q' \ne q,\\n'=\{1,\dots,N\}, k'=\{1,\dots,K\}}} {\bf h}_{q'n',qk}^H {\bf w}_{q'n'k'} s_{q'k'}}_{\substack{\text{inter-cluster}\text{ interference}}}
+ z_{qk}
\end{align}
where $s_{qk}\in\mathbb{C}$ is the power-limited signal sent to user $k$ by the serving RRHs in cell $q$, that is $\mathbb{E}\{{\bf s}_q {\bf s}_q^H\} = p {\bf I}_K$ for ${\bf s}_q=[s_{q1},\dots,s_{qK}]$, with $p$ being the total power budget for the RRHs in the cell, which is referred as vector normalization~\cite{6823643}.

The vector ${\bf h}_{qn,qk}\in\mathbb{C}^{M}$ denotes the channel between RRH $n$ in cell $q$ and user $k$ in cell $q$ and accounts for the small-scale and large-scale fading. We have ${\bf h}_{qn,qk} = \sqrt{\ell_k(x_{qn},y_{qn})}{\bf g}_{qn,qk}$, where ${\bf g}_{qn,qk}\sim\mathcal{CN}(0,{\bf I}_M)$ is a white complex Gaussian random vector representing the Rayleigh fading, while $\ell_k(x_{qn},y_{qn})=\left(1+d_{qn,qk}/d_0\right)^{-\alpha}$ is the path loss of the signal, with $d_{qn,qk} =\sqrt{(x_{qn} - \widetilde{x}_k)^2 + (y_{qn} - \widetilde{y}_k)^2}$ being the distance between RRH $n$ and user $k$, $d_0$ is the reference distance, and $\alpha$ is the path loss exponent. The terms $x_{qn}$ and $ y_{qn}$ are the $x,y$-coordinates of RRH $n$ and will be our optimization variables. 
Furthermore, ${\bf w}_{qnk}\in\mathbb{C}^M$ is the linear precoding vector used by RRH $n$ to serve user $k$ in its cell $q$, and $z_{qk}\sim \mathcal{CN}(0,\sigma_z^2)$ is the independent additive white Gaussian noise (AWGN), with $\sigma^2_z$ being the noise power.

The RRHs use ZF-BF to serve the $K$ users in each cell. To construct the beamformer, we denote the concatenation of the channel matrix for all the users in cell $q$ as ${\bf H}_{q}=[{\bf h}_{q1}\dots{\bf h}_{qK}]\in\mathbb{C}^{NM\times K}$ with presumably linearly independent rows, with ${\bf h}_{qk}=[{\bf h}_{qk,qk}^T\dots{\bf h}_{qN,qk}^T]^T\in\mathbb{C}^{NM}$ being the concatenation of the access channels for user $k$. We can write ${\bf h}_{qk}={\bf D}_{qk}^{1/2}{\bf g}_{qk}$, where ${\bf D}_{qk}\in \mathbb{C}^{NM\times NM}$ is a diagonal matrix representing the large-scale fading between all the RRHs in the network and user $k$. Hence, 
$[{\bf D}_{qk}]_{mm}=\ell_k(x_{qn},y_{qn})$ where $n=\lceil{m/M}\rceil$ for $m=\{1,\dots,NM\}$. Furthermore, ${\bf g}_{qk}\in \mathbb{C}^{NM}$ represents the vector of the small-scale fading coefficients between the antennas of the RRHs and user $k$.

The precoding matrix ${\bf W}_{q}$ is designed for each cell $q$ assuming CSI at the RRHs is available. We have ${\bf W}_{q}=[{\bf v}_{q1} \dots {\bf v}_{qK}]\in\mathbb{C}^{NM\times K}$ with ${\bf v}_{qk}=[{\bf w}_{q1k}^T \dots {\bf w}_{qNk}^T]^T\in\mathbb{C}^{NM}$, where ${\bf w}_{qnk} \in \mathbb{C}^N$ is the precoding vector used at RRH $n$, and is given by
\begin{align}\label{eq:beamformerAccess}
{\bf W}_{q}={\bf \widetilde{W}}_{q}\bm{\mu}_q=\left({\bf H}_{q}\right)^\dagger \bm{\mu}_q = {\bf H}_{q}\left({\bf H}_{q}^H{\bf H}_{q}\right)^{-1} \bm{\mu}_q
\end{align}
Here, $\bm{\mu}_q \in \mathbb{C}^{K \times K}$ is a diagonal matrix which satisfies the power budget $\mathbb{E}\left\{\text{tr}\left\{ {\bf W}_{q} {\bf W}_{q}^H\right\}\right\} = p$, i.e., using average power normalization~\cite{ZFBFLetter9141340}, and its $k^{th}$ entry is
\begin{align}
\mu_{qk} = \left[\bm{\mu}_q\right]_{k,k} = \sqrt{p K^{-1} \mathbb{E}\{\|{\bf \widetilde{v}}_{qk}\|^2\}^{-1}}
\end{align}
with ${\bf \widetilde{v}}_{qk}=[{\bf H}_{q}\left({\bf H}_{q}^H{\bf H}_{q}\right)^{-1}]_{.k}$ is $k^{th}$ column of the matrix.
\subsection{Backhaul Transmission Scheme}
We consider point-to-point single-input single-output (SISO) noise-limited wireless transmission between the CUs and their RRHs (representing an upper-bound for the performance of a SISO solution). We assume that the CU implements directional antennas, providing parallel \emph{Rician channels} between the CU and the $N$ RRHs. 

The downlink average achievable rate between CU $q$ and its served RRHs is defined as
\begin{align}
R_{qn}^{(b)}\left(x_{qn},y_{qn}\right)&=\mathbb{E}\left\{\log\left(1+\rho_c\left|\bar{g}_{qn}\right|^2 \bar{\ell}_q(x_{qn},y_{qn}) \right)\right\}
\end{align}
where $\rho_c=\frac{p_c}{\sigma^2}$, $\bar{\ell}_q(x_{qn},y_{qn})=\left(1+\bar{d}_{qn}/d_0\right)^{-\alpha}$ is the path loss on the backhaul link, and it depends on the distance between CU $q$ and RRH $n$ denoted as $\bar{d}_{qn}=\sqrt{( x_{qn} - \bar{x}_q )^2 + ( y_{qn} - \bar{y}_q )^2}$, where $\left(\bar{x}_q, \bar{y}_q \right)$ are the coordinates of the CU. The term $\bar{g}_{qn}$ is the small-scale fading of the channel, which follows a Rician fading model. This Rician channel is characterized by the Rician parameter (the Rician K-factor) $\mathcal{K}=\eta_1^2/\eta_2^2$, where $\eta_1^2$ and $\eta_2^2$ represent the power of the line-of-sight (LoS) and NLoS components respectively.

%
%
\section{Problem Formulation}\label{sec:problemFormulation}
We aim to find the optimal deployment for the RRH locations in the presence of specific traffic distributions, inter-cell interference, and most importantly, under limited capacity wireless backhaul. These important factors will be characterized in the next subsections.
\subsection{Traffic Distribution}\label{sec:traffic_dist}
We use a traffic probability density function (PDF) that is a combination of both a uniform distribution, chosen with probability $P_0$, and a number $N_h$ of bivariate normal distributions, representing hotspots (a choice of small $P_0$ means more dense hotspots). The number of these hotspots is uniformly distributed, i.e., $N_h \sim U\left(N_h^\text{min}, N_h^\text{max} \right)$. This number can be generated for the whole network or for each cell separately. These hotspots are centered at locations $\left[{\bf \widetilde{x}}_h,{\bf \widetilde{y}}_h\right] = \left[\left[{\widetilde{x}}_{h1},\dots,{\widetilde{x}}_{h N_h}\right]^T, \left[{\widetilde{y}}_{h1}, \dots, {\widetilde{y}}_{h N_h}\right]^T\right] \in \mathbb{R}^{N_h\times 2}$, and they have equal variances $\sigma_h^2$ on both the $x$ and $y$ axis, without any correlation between the two axis. Therefore, the cell traffic distribution PDF bounded by the cell boundary $\mathcal{B}_{q}$ is defined as
\begin{align}\label{eq:traffic_dist}
&f_q(\widetilde{x}_k,\widetilde{y}_k) = f_0 \left(P_0 \left( \frac{1}{B_{q}} \right)
+ \left( 1 - P_0 \right)
\right.
\nonumber \\
&\ 
\resizebox{0.42\textwidth}{!}
{$ \displaystyle
\left.
\frac{1}{N_h 2 \pi \sigma_h^2} \sum_{i \in 1}^{N_h} \left( \exp\left(-\frac{\left(\widetilde{x}_k - \widetilde{x}_{h_i}\right)^2+\left(\widetilde{y}_k - \widetilde{y}_{h_i}\right)^2}{2\sigma_h^2}\right) \right) \right)
$}
\end{align}
where $B_{q}$ (different from cell boundary $\mathcal{B}_{q}$) is the area of cell $q$, $f_0$ is a normalizing factor that can be calculated numerically to normalize the PDF. This traffic model is flexible in the sense that it can be constructed from a traffic survey that identifies the locations of hotspots in the network.

\subsection{Access Channel Spectral Efficiency}
On the access channel, the ZF beamformer is formed per cell $q$. Hence, the intra-cluster interference found in~\eqref{eq:signalModel_accessChannel} will be completely canceled. 
As a result, the mean achievable spectral efficiency over the access channel for user $k$ in cell $q$ is
\vspace{-0.5em}
\begin{align}
R&_{qk}^{(a)}\left({\bf x},{\bf y}\right)=
\nonumber \\
&\resizebox{0.47\textwidth}{!}
{$ \displaystyle
\mathbb{E}\left\{\log\left(1+\frac{ 
\mu_{qk}^2
} {
\sum_{\substack{n'=\{1,\dots,N\}, q' \ne q,\\ k'=\{1,\dots,K\}}}\mid {\bf h}_{q'n',qk}^H {\bf w}_{q'n'k'}\mid^2 + \sigma_z^2
} \right)\right\}
$} 
\end{align}

The concatenated form of the RRHs' locations in the network is written as $\left[{\bf x},{\bf y}\right]$, where the the x-coordinates are ${\bf x}=[{\bf x}_1^T,\dots,{\bf x}_q^T,\dots,\dots,{\bf x}_Q^T]^T \in \mathbb{C}^{NQ}$, for $q \in \{1,\dots,Q\}$ (similarly for ${\bf y}$). These will form our optimization variables. We use $[{\bf x}_q, {\bf y}_q]$ to refer to the locations of the RRHs in cell $q$ and $[{\bf x}_{-q},{\bf y}_{-q}]$ to denote the locations of all the RRHs in the network except those in cell $q$. 

The spectral efficiency over the access channels depends on the distances between all the RRHs found in the network and the user $k$, which in its turn depends on the locations of all the RRHs. Given the distances to the serving and interfering RRHs, the lower-bound of spectral efficiency can be written as~\cite{8529184}
\vspace{-0.6em}
\begin{align}
R_{qk}^{(a)}\left({\bf x}, {\bf y}\right)=
\resizebox{0.37\textwidth}{!}
{$ \displaystyle
	\log\left(1+\displaystyle \gamma_k\left({\bf x}_{-q},{\bf y}_{-q}\right)^{-1}\sum_{n=1}^{N}\ell_k(x_{qn},y_{qn}) \right)
	$}
\end{align}
\vspace{-1.3em}
$\!\!$with
\begin{align}
\gamma_k\left({\bf x}_{-q},{\bf y}_{-q}\right)=
\resizebox{0.35\textwidth}{!}
{$ \displaystyle
	\frac{NK}{(NM-K)\rho}\left(\displaystyle \frac{M \rho}{K} \sum_{q' = 1,q'\ne q}^{Q} \text{ICI}_{q'k}({\bf x}_{q'},{\bf y}_{q'}) + 1\right)
	$}
\end{align}
\vspace{-1.5em}
\begin{align}
\text{ICI}_{q'k}({\bf x}_{q'},{\bf y}_{q'})=\sum_{l = 1}^{N}\ell_k(x_{q'l},y_{q'l})\sum_{j=1}^{K}\frac{\ell_j(x_{q'l},y_{q'l})}{\xi(q',j)}
\end{align}
where $\rho=\frac{p}{\sigma_z^2}$, and $\text{ICI}_{q'k}({\bf x}_{q'},{\bf y}_{q'})$ denotes the inter-cell interference (ICI) contributed by cell $q'$ to the user under test and $\xi(q',j)=\text{tr}\{{\bf D}_{q'j}\}=M\sum_{m=1}^{N}\ell_j(x_{q'm},y_{q'm})$. As can be seen, this interference depends on the locations of the users in the interfering cells because the beamformers in each cell is designed based on the channels of the users in these cells.

Using the traffic distribution PDF, $f_q(\widetilde{x}_k,\widetilde{y}_k)$, in~\eqref{eq:traffic_dist}, we further average this achievable rate over $f_{q'}(\widetilde{x}_k,\widetilde{y}_k)$ in the interfering cells by writing the ICI term as
\vspace{-0.5em}
\begin{align}
&\text{ICI}_{q'k}({\bf x}_{q'},{\bf y}_{q'}) =
\resizebox{0.33\textwidth}{!}
{$ \displaystyle
	\sum_{l=1}^{N}\ell_k(x_{q'l},y_{q'l})K\mathbb{E}_{\widetilde{x}_j,\widetilde{y}_j}\left\{\frac{\ell_j(x_{q'l},y_{q'l})}{\xi(q',j)}\right\}
	$}
\nonumber\\
&\ 
=
\resizebox{0.44\textwidth}{!}
{$ \displaystyle
	\sum_{l = 1}^{N}\ell_k(x_{q'l},y_{q'l}) K
	\int \!\!  \int_{\widetilde{x}_j,\widetilde{y}_j\in \mathcal{B}_{q'}} \!\!
	\frac{\ell_j(x_{q'l},y_{q'l})}{\xi(q',j)} f_{q'}(\widetilde{x}_j,\widetilde{y}_j) \diff \widetilde{x}_j \diff \widetilde{y}_j
	$}
\end{align}
where, as indicated earlier, $\mathcal{B}_{q'}$ is the boundary of cell $q'$.
%
%
\subsection{Problem Definition}
\par We define our problem of optimizing the locations of the RRHs in the network as
\begin{subequations}\label{eq:opt1Formulation}
	\begin{align}
	&\underset{{\bf x},{\bf y}}{\text{max}}\quad \mathbb{E}_{\widetilde{x}_k,\widetilde{y}_k}\left\{ R_{qk}^{(a)}({\bf x},{\bf y}) \right\},\ \forall q
	\label{eq:ObjectiveFunct1}\\
	&\text{s.t.}\quad
	\resizebox{0.44\textwidth}{!}
	{$\displaystyle
		\displaystyle \mathbb{P}\left\{\omega_c R_{qn}^{(b)}(x_{qn},y_{qn}) \le K \omega
		\mathbb{E}_{\widetilde{x}_k,\widetilde{y}_k}\left\{ R_{qk}^{(a)}({\bf x},{\bf y}) \right\}
		\right\}\le\epsilon,
		$}
	\nonumber \\
	& \quad \quad \quad n=1,\dots,N
	\label{eq:Opt1_v2_Constraint1}
	\end{align}
\end{subequations}
where
\begin{align} \label{eq:trafficExpectation}
\mathbb{E}_{\widetilde{x}_k,\widetilde{y}_k}\left\{ R_{qk}^{(a)}({\bf x},{\bf y}) \right\} = 
\resizebox{0.29\textwidth}{!}
{$\displaystyle
	\int \!\!  \int_{\widetilde{x}_k,\widetilde{y}_k\in \mathcal{B}_{q}} R_{qk}^{(a)}({\bf x},{\bf y}) f_q(\widetilde{x}_k,\widetilde{y}_k) \diff \widetilde{x}_k \diff \widetilde{y}_k
	$}
\end{align}
%
The terms $\omega$, $\omega_c$ are the bandwidth allocated for the access channel and the backhaul, respectively, and $\epsilon$ is the allowed backhaul outage probability for each CU-RRH link.
\par The formulation in~\eqref{eq:opt1Formulation} maximizes the average spectral efficiency of \textit{typical user} $k$ in the network, and this spectral efficiency is averaged over the traffic distribution in the user's cell as shown in~\eqref{eq:ObjectiveFunct1}, which means it maximizes the spectral efficiency for the system. Additionally, the $N$ constraints in~\eqref{eq:Opt1_v2_Constraint1} place an upper bound on the sum of the rates over the access channel with respect to the backhaul achieved capacity. If this constraint is not respected the backhaul will experience an outage.

\noindent \emph{{Proposition}:} The probability of outage, $P_{qn}\left({\bf x},{\bf y}\right)$ is given by
\begin{align}\label{eq:P_qn}
P_{qn}\left({\bf x},{\bf y}\right)&=
\resizebox{0.38\textwidth}{!}
{$\displaystyle
	\mathbb{P}\left\{ \omega_c R_{qn}^{(b)}(x_{qn},y_{qn}) \le K \omega 
	\mathbb{E}_{\widetilde{x}_k,\widetilde{y}_k}\left\{ R_{qk}^{(a)}({\bf x},{\bf y}) \right\}
	\right\}
	$}
\nonumber \\
&=1-Q_1\left(\frac{\sqrt{2}\eta_1}{\eta_2},\frac{\sqrt{2 \zeta_{qn}({\bf x},{\bf y})}}{\eta_2}\right)
\end{align}
where $Q_1(.)$ is the Marcum $Q$-function, and
\begin{align}
\resizebox{0.48\textwidth}{!}
{$\displaystyle
	\zeta_{qn}({\bf x},{\bf y})=\frac{1}{\rho_c \bar{\ell}_q(x_{qn},y_{qn})}\left(
	\exp\left(K \frac{\omega}{\omega_c} 
	\mathbb{E}_{\widetilde{x}_k,\widetilde{y}_k}\left\{ R_{qk}^{(a)}({\bf x},{\bf y}) \right\}
	\right)
	-1\right)
	$}
\end{align}
\begin{proof}
	Please see Appendix~\ref{appendix:RateConstraint}.
\end{proof}
\section{Optimizing RRH Placement Under Backhaul Constraints}\label{sec:RRHsPlacement}
In the next subsections, we solve the optimization problem in~\eqref{eq:opt1Formulation} using two different approaches.
\subsection{Direct Approach}
For notational simplicity we define the following term.
\begin{align}
&J_{1_{qn}}({\bf x},{\bf y}) = \frac{\sqrt{2 \zeta_{qn}({\bf x},{\bf y})}}{\eta_2} = J_{2_{qn}}(x_{qn},y_{qn}) \times J_{3_{q}}({\bf x},{\bf y})
\nonumber \\
&
=
\resizebox{0.45\textwidth}{!}
{$ \displaystyle
	\frac{1}{\eta_2} \sqrt{\frac{2}{\rho_c \bar{\ell}_q(x_{qn},y_{qn})}}
	\times \sqrt{\left(\exp\left( K \frac{\omega}{\omega_c} \mathbb{E}_{\widetilde{x}_k,\widetilde{y}_k}\left\{ R_{qk}^{(a)}({\bf x},{\bf y}) \right\}\right)
		-1\right)}
	$}
\end{align}
We can write the Lagrangian formulation of our problem as
\begin{align}\label{eq:Lagrangian_firstScheme}
\mathcal{L}({\bf x},{\bf y},\bm{\lambda})=
\resizebox{0.37\textwidth}{!}
{$ \displaystyle
- \mathbb{E}_{\widetilde{x}_k,\widetilde{y}_k}\left\{ R_{qk}^{(a)}({\bf x},{\bf y}) \right\} 
+ \sum_{n=1}^{N}\lambda_n\left(P_{qn}\left({\bf x},{\bf y}\right)-\epsilon\right)
$}
\end{align}
where the $\bm{\lambda} \succcurlyeq 0$ denotes the vector of Lagrange multipliers. For a specific RRH $m\in\{1,\dots,N\}$ in cell $q$, we differentiate the Lagrangian formulation with respect to the x-coordinate $x_{qm}$ of RRH $m$, set it to zero, and obtain an iterative formula for $x_{qm}$ that can be written as
\begin{align}\label{eq:xmUpdate}
x_{qm}^{(i+1)} = 
\resizebox{0.41\textwidth}{!}
{$\displaystyle
	\frac{ \displaystyle \left(1 - A_{2} - A_{4}\right) \int \!\!  \int_{\widetilde{x}_k,\widetilde{y}_k\in \mathcal{B}_{q}} \widetilde{x}_k A_{1}(\widetilde{x}_k,\widetilde{y}_k) f_q(\widetilde{x}_k,\widetilde{y}_k) \diff \widetilde{x}_k \diff \widetilde{y}_k
		+ A_{3} \bar{x}_q }
	{ \displaystyle \left(1 - A_{2} - A_{4}\right) \int \!\!  \int_{\widetilde{x}_k,\widetilde{y}_k\in \mathcal{B}_{q}} A_{1}(\widetilde{x}_k,\widetilde{y}_k) f_q(\widetilde{x}_k,\widetilde{y}_k) \diff \widetilde{x}_k \diff \widetilde{y}_k
		+ A_{3}}
	$}
\end{align}
The same formulation applies by differentiating the Lagrangian with respect to the y-coordinate.
\begin{align}\label{eq:ymUpdate}
y_{qm}^{(i+1)} =
\resizebox{0.41\textwidth}{!}
{$\displaystyle
	\frac{ \displaystyle \left(1 - A_{2} - A_{4}\right) \int \!\!  \int_{\widetilde{x}_k,\widetilde{y}_k\in \mathcal{B}_{q}} \widetilde{y}_k A_{1}(\widetilde{x}_k,\widetilde{y}_k) f_q(\widetilde{x}_k,\widetilde{y}_k) \diff \widetilde{x}_k \diff \widetilde{y}_k
		+ A_{3} \bar{y}_q }
	{ \displaystyle \left(1 - A_{2} - A_{4}\right) \int \!\!  \int_{\widetilde{x}_k,\widetilde{y}_k\in \mathcal{B}_{q}} A_{1}(\widetilde{x}_k,\widetilde{y}_k) f_q(\widetilde{x}_k,\widetilde{y}_k) \diff \widetilde{x}_k \diff \widetilde{y}_k
		+ A_{3} }
	$}
\end{align}
where
\begin{align}
A_{1}(\widetilde{x}_k,\widetilde{y}_k) =
\resizebox{0.37\textwidth}{!}
{$\displaystyle
	\frac{ \gamma_k\left({\bf x}_{-q},{\bf y}_{-q}\right)^{-1} \left(1+d_{qm,qk}/d_0\right)^{-1-\alpha}}{ d_{qm,qk}\left(1+\gamma_k\left({\bf x}_{-q},{\bf y}_{-q}\right)^{-1} \displaystyle \sum_{n=1}^{N}\left(1+d_{qn,qk}/d_0\right)^{-\alpha}\right)}
	$}
\end{align}
\begin{align}
&A_{2} =
\frac{ K\frac{\omega}{\omega_c} \lambda_m J_{1_{qm}}({\bf x},{\bf y}) \exp\left(K \frac{\omega}{\omega_c} \mathbb{E}_{\widetilde{x}_k,\widetilde{y}_k}\left\{ R_{qk}^{(a)}({\bf x},{\bf y}) \right\}\right) }{\eta_2 \sqrt{2 \rho_c \left(\exp\left(K \frac{\omega}{\omega_c} \mathbb{E}_{\widetilde{x}_k,\widetilde{y}_k}\left\{ R_{qk}^{(a)}({\bf x},{\bf y}) \right\}\right) - 1\right)} }
\nonumber \\
&
\times
\resizebox{0.465\textwidth}{!}
{$\displaystyle
	\exp\left(-\left(\frac{\eta_1^2}{\eta_2^2}+\frac{\left(J_{1_{qm}}({\bf x},{\bf y})\right)^2}{2}\right)\right)
	{}_0F_1\left(;1;\frac{\eta_1^2}{2\eta_2^2}\left(J_{1_{qm}}({\bf x},{\bf y})\right)^2 \right) 
	\left(1 + \frac{\bar{d}_{qm}}{d_0}\right)^{\frac{\alpha}{2}}
	$}
\end{align}
\begin{align}
&A_{3} =
\resizebox{0.43\textwidth}{!}
{$\displaystyle
	\frac{ \lambda_m J_{1_{qm}}({\bf x},{\bf y}) \sqrt{\left(\exp\left(K \frac{\omega}{\omega_c} 
			\mathbb{E}_{\widetilde{x}_k,\widetilde{y}_k}\left\{ R_{qk}^{(a)}({\bf x},{\bf y}) \right\}
			\right) - 1\right)} }{ \eta_2 \sqrt{2 \rho_c } }
	$}
\nonumber \\
&
\times
\resizebox{0.46\textwidth}{!}
{$\displaystyle
	\exp\left(-\left(\frac{\eta_1^2}{\eta_2^2}+\frac{\left(J_{1_{qm}}({\bf x},{\bf y})\right)^2}{2}\right)\right)
	{}_0F_1\left(;1;\frac{\eta_1^2}{2\eta_2^2}\left(J_{1_{qm}}({\bf x},{\bf y})\right)^2 \right)
	\left( \frac{ \left(1 + \bar{d}_{qm}/d_0\right)^{\frac{\alpha}{2}-1} } {\bar{d}_{qm} } \right)
	$}
\end{align}
\begin{align}
&A_{4} =
\resizebox{0.43\textwidth}{!}
{$\displaystyle
	\frac{ K \frac{\omega}{\omega_c} \exp\left(K \frac{\omega}{\omega_c} 
		\exp\left(-\left(\frac{\eta_1^2}{\eta_2^2}+\frac{\left(J_{1_{qm}}({\bf x},{\bf y})\right)^2}{2}\right)\right)
		\right) }{\eta_2 \sqrt{2 \rho_c \left(\exp\left(K \frac{\omega}{\omega_c}
			\exp\left(-\left(\frac{\eta_1^2}{\eta_2^2}+\frac{\left(J_{1_{qm}}({\bf x},{\bf y})\right)^2}{2}\right)\right)
			\right) - 1\right)} }
	$}
\nonumber\\
&\times\!\!
\sum_{n \in \mathcal{D}_q, n \ne m}
\lambda_n J_{1_{qn}}({\bf x},{\bf y})	
\exp\left(-\left(\frac{\eta_1^2}{\eta_2^2}+\frac{\left(J_{1_{qn}}({\bf x},{\bf y})\right)^2}{2}\right)\right)
\nonumber \\
&
\quad \quad
\times
{}_0F_1\left(;1;\frac{\eta_1^2}{2\eta_2^2}\left(J_{1_{qn}}({\bf x},{\bf y})\right)^2 \right)
\left(1 + \frac{\bar{d}_{qn}}{d_0}\right)^{\frac{\alpha}{2}}
\end{align}
\begin{proof}
	Follows from the derivative chain rule and has been skipped due to lack of space.
\end{proof}
The terms $A_{1}(\widetilde{x}_k,\widetilde{y}_k), A_{2}, A_{3}$ and $A_{4}$ depend on the RRHs locations, but we do not write the RRHs ${\bf x}$ and ${\bf y}$ coordinates as parameters to minimize the notation. Here ${}_0F_1(;.;.)$ is the regularized confluent Hypergeometric function, and we can write it in an alternate form as a function of the modified Bessel function of first kind as ${}_0F_1\left(;1;\frac{\eta_1^2}{2\eta_2^2}\left(J_{1_{qm}}({\bf x},{\bf y})\right)^2 \right)=I_0\left( \frac{\sqrt{2}\eta_1}{\eta_2}J_{1_{qm}}({\bf x},{\bf y}) \right)$, i.e., ${}_0F_1\left(;1;z \right)=I_0\left( 2\sqrt{z} \right)$.
%
%
%
\par The derivative of $\mathcal{L}({\bf x},{\bf y},\bm{\lambda})$ in~\eqref{eq:Lagrangian_firstScheme} with respect to $\lambda_m$ is
\begin{align}
&\mathcal{L}_{\lambda_m}({\bf x},{\bf y},\bm{\lambda}) =
\frac{\partial \mathcal{L}({\bf x},{\bf y},\bm{\lambda})}{\partial \lambda_m}
\nonumber \\
&\quad\quad
= \left(1-Q_1\left(\frac{\sqrt{2}\eta_1}{\eta_2},\frac{\sqrt{2 \zeta_{qm}({\bf x},{\bf y})}}{\eta_2}\right)-\epsilon \right)
\end{align}
Using the batch gradient descent, we can obtain an iterative formula for $\lambda^{(i+1)}_m$ as
\begin{align}\label{eq:lambdaUpdate}
\lambda^{(i+1)}_m = \left[\lambda^{(i)}_m + \nu \mathcal{L}_{\lambda_m}({\bf x}^{(i)},{\bf y}^{(i)},\bm{\lambda}^{(i)}) \right]^+
\end{align}
where $\left[\cdot\right]^+=\max\left(\cdot,0\right)$, and $\nu \in \mathbb{R}_+$ is a step size chosen small enough to guarantee convergence. Based on this analysis, we can construct Algorithm~\ref{algortihm:iterativeAlgo} to obtain the optimal locations of the RRHs in the network as described below.
\begin{algorithm}[h!]
	\SetAlgoLined
	\SetInd{0.1em}{1em}
	\caption{RRHs Locations Optimization}
	\label{algortihm:iterativeAlgo}
	\small
	Generate random locations for the RRHs in all cells \label{algo:randomLoc}\\
	Define $d_\text{max}$ big enough\\
	\While{$d_{\text{max}} > d_\text{cvg}$}{
		\For{ $q \in \{1,\dots,Q\}$}{
			Wrap-around cells to make cell $q$ at center \label{algo:wraparound}\\
			\For{ $m \in \mathcal{D}_q$}{
				Update $x_{qm}$, $y_{qm}$, $\lambda_m$ using eq.~\eqref{eq:xmUpdate},~\eqref{eq:ymUpdate},~\eqref{eq:lambdaUpdate} \label{algo:updateRRHsLoc}\\
			}
			$d_q=\underset{n}{\max}\left\{\underset{x,y}{\max}\left\{|x_{qn}^{(i+1)} - x_{qn}^{(i)}|, |y_{qn}^{(i+1)} - y_{qn}^{(i)}|\right\}\right\}$
		}
		$d_\text{max} = \underset{q}\max\ d_q$ \label{algo:d_max}
	}
\end{algorithm}
\par The algorithm starts by choosing random locations for the RRHs (Step \ref{algo:randomLoc}). Then, we define a $d_\text{max}$ as a distance large enough to start the locations update. For the RRHs location update in each algorithm iteration, we perform a cell wrap-around (Step \ref{algo:wraparound}) to place the cell of these RRHs at the center, hence eliminating network border effect. After that, we update the locations of the RRHs in this cell using the indicated equations in Step \ref{algo:updateRRHsLoc}, and we calculate the maximum RRHs location change. We do one iteration for each cell at a time until we iterate through all the network cells. Hence, the total number of iterations will be the same for all the cells, and the updated RRHs locations in the interfering cells will be used, which is very reliable. At last, the algorithm convergence is determined when the maximum change in the RRHs locations ($d_\text{max}$ in Step \ref{algo:d_max}) is lower than a small distance $d_\text{cvg}$. We note that our scheme is sub-optimal mainly because of the complex traffic distribution shown in~\eqref{eq:traffic_dist}.
\subsection{Distance-based Approach}
\par Choosing an appropriate step size $\nu$ for ${\bf \lambda}$ update in~\eqref{eq:lambdaUpdate} can be tricky especially when the constraint~\eqref{eq:Opt1_v2_Constraint1} is tight. To address this issue, we define an equivalent approach that is distance-based to solve problem~\eqref{eq:opt1Formulation} by formulating it as
\begin{subequations}
	\begin{align}
	&\underset{{\bf x},{\bf y}}{\text{max}}\quad 
	\mathbb{E}_{\widetilde{x}_k,\widetilde{y}_k}\left\{ R_{qk}^{(a)}({\bf x},{\bf y}) \right\}, \forall q
	\label{eq:ObjectiveFunct1_v3}\\
	&\text{s.t.}\quad
	\frac{\bar{d}_{qn}} {\bar{d}_\text{out}} \le 1,
	\quad n=1,\dots,N
	\label{eq:Opt1_v3_Constraint1}
	\end{align}
\end{subequations}
where $\bar{d}_\text{out}$ is the maximum allowed backhaul distance to guarantee that the outage is lower than $\epsilon$. When the achievable spectral efficiency over the access channel is fixed, $\bar{d}_\text{out}$ can be easily obtained using a bisection search 
to obtain $P_{qn}\left({\bf x},{\bf y}\right) = \epsilon$ defined in~\eqref{eq:P_qn}.
%
Once $\bar{d}_\text{out}$ is found, we can write an iterative formula for $x_{qm}$ and $y_{qm}$ as
\begin{align}\label{eq:xmUpdate2}
\resizebox{0.48\textwidth}{!}
{$\displaystyle
	x_{qm}^{(i+1)} = \frac{ \frac{\alpha}{d_0} \displaystyle \int \!\!  \int_{\widetilde{x}_k,\widetilde{y}_k\in \mathcal{B}_{q}} \widetilde{x}_k A_{1}(\widetilde{x}_k,\widetilde{y}_k) f_q(\widetilde{x}_k,\widetilde{y}_k) \diff \widetilde{x}_k \diff \widetilde{y}_k
		+ \frac{\lambda_m}{\bar{d}_\text{out}\bar{d}_{qm}} \bar{x}_q }
	{ \frac{\alpha}{d_0} \displaystyle \int \!\!  \int_{\widetilde{x}_k,\widetilde{y}_k\in \mathcal{B}_{q}} A_{1}(\widetilde{x}_k,\widetilde{y}_k) f_q(\widetilde{x}_k,\widetilde{y}_k) \diff \widetilde{x}_k \diff \widetilde{y}_k
		+ \frac{\lambda_m}{\bar{d}_\text{out}\bar{d}_{qm}}}
	$}
\end{align}
Similarly for the y-coordinates:
\begin{align}\label{eq:ymUpdate2}
\resizebox{0.48\textwidth}{!}
{$\displaystyle
	y_m^{(i+1)} = \frac{ \frac{\alpha}{d_0} \displaystyle \int \!\!  \int_{\widetilde{x}_k,\widetilde{y}_k\in \mathcal{B}_{q}} \widetilde{y}_k A_{1}(\widetilde{x}_k,\widetilde{y}_k) f_q(\widetilde{x}_k,\widetilde{y}_k) \diff \widetilde{x}_k \diff \widetilde{y}_k
		+ \frac{\lambda_m}{\bar{d}_\text{out}\bar{d}_{qm}} \bar{y}_q }
	{ \frac{\alpha}{d_0} \displaystyle \int \!\!  \int_{\widetilde{x}_k,\widetilde{y}_k\in \mathcal{B}_{q}} A_{1}(\widetilde{x}_k,\widetilde{y}_k) f_q(\widetilde{x}_k,\widetilde{y}_k) \diff \widetilde{x}_k \diff \widetilde{y}_k
		+ \frac{\lambda_m}{\bar{d}_\text{out}\bar{d}_{qm}} }
	$}
\end{align}
%
%
%
To update $\lambda_m^{i}$, we perform another bisection search, such that $\bar{d}_{qm} = \bar{d}_\text{out}$. This method eliminates the need for a step size $\nu$ to update $\bm{\lambda}$, and at the same time~\eqref{eq:xmUpdate2} and~\eqref{eq:ymUpdate2} guarantee that the RRHs will be placed in the locations that maximize~\eqref{eq:ObjectiveFunct1_v3} as we will see in the results section. Consequently, we can obtain the optimal locations of the RRHs in our network using Algorithm~\ref{algortihm:iterativeAlgo}, but with replacing Step~\ref{algo:updateRRHsLoc} with two steps; one that obtains $\bar{d}_\text{out}$ using bisection search and the other updates $x_{qm}$ and $y_{qm}$ using~\eqref{eq:xmUpdate2} and~\eqref{eq:ymUpdate2} respectively.

\section{Numerical Results}\label{sec:results}
We consider a network of $Q=9$ cells with wrap-around and cell dimension of $1000 \times 1000$ meters. The cells have square shapes, but any other preferred shape can be used if needed, e.g., circular. Moreover, we consider a system of $25$ resource blocks (RBs), where each RB has a bandwidth (BW) of $180$ KHz. We use the traffic distribution in~\eqref{eq:traffic_dist} to model the locations of users. We summarize the rest of the simulation parameters in Table \ref{table:sim_parameters}, where the available system bandwidth is divided between the backhaul (of BW $\omega_c$) and the access channel (of BW $\omega$).
\begin{table}[t]
	\centering
	\begin{tabular}{|p{0.22\linewidth}|p{0.26\linewidth}|p{0.36\linewidth}|}
		\hline
		\hline
		& \multicolumn{1}{l|}{ \textit{\textbf{Parameter}}} & \multicolumn{1}{l|}{\textit{\textbf{Value}}}\\
		\hline
		Cell config. & $Q$, $N$, $M$, $K$ & $9$, $10$, $8$, $10$\\
		\hline
		Power & $p$, $p_c$  & $30$ dBm, $45$ dBm\\
		\hline
		Bandwidth &RB, $\omega$, $\omega_c$ & $180$ KHz, $5$ RBs, $20$ RBs\\
		\hline
		Noise & spectral density $S_z$, noise figure $F_z$ & $-174$ dBm/Hz, $8$ dBm\\
		\hline
		Hotspots & \pbox{20cm}{$P_0$, $\sigma_h$;\\ $N_h^\text{min}$, $N_h^\text{max}$} & \pbox{20cm}{$0.1$, $100$ meters;\\ Per network: $2Q$, $4Q$} \\
		\hline
		Path loss, Fading &$d_0$, $\alpha$, $\mathcal{K}$, $\eta_1$, $\eta_2$&$0.392$ meters, $3.76$, $15$ dB, $8$, $\sqrt{2}$\\
		\hline
		Algorithm &$\epsilon$, $d_\text{cvg}$, $\nu$ & $0.2$, $1$ meter, $1$\\
		\hline
		\hline
	\end{tabular}
	\caption{Simulation parameters.}
	\label{table:sim_parameters} 
	\vspace{-1em}  
\end{table}
\begin{figure}[t]
	\centering
	\includegraphics[width=0.4\textwidth]{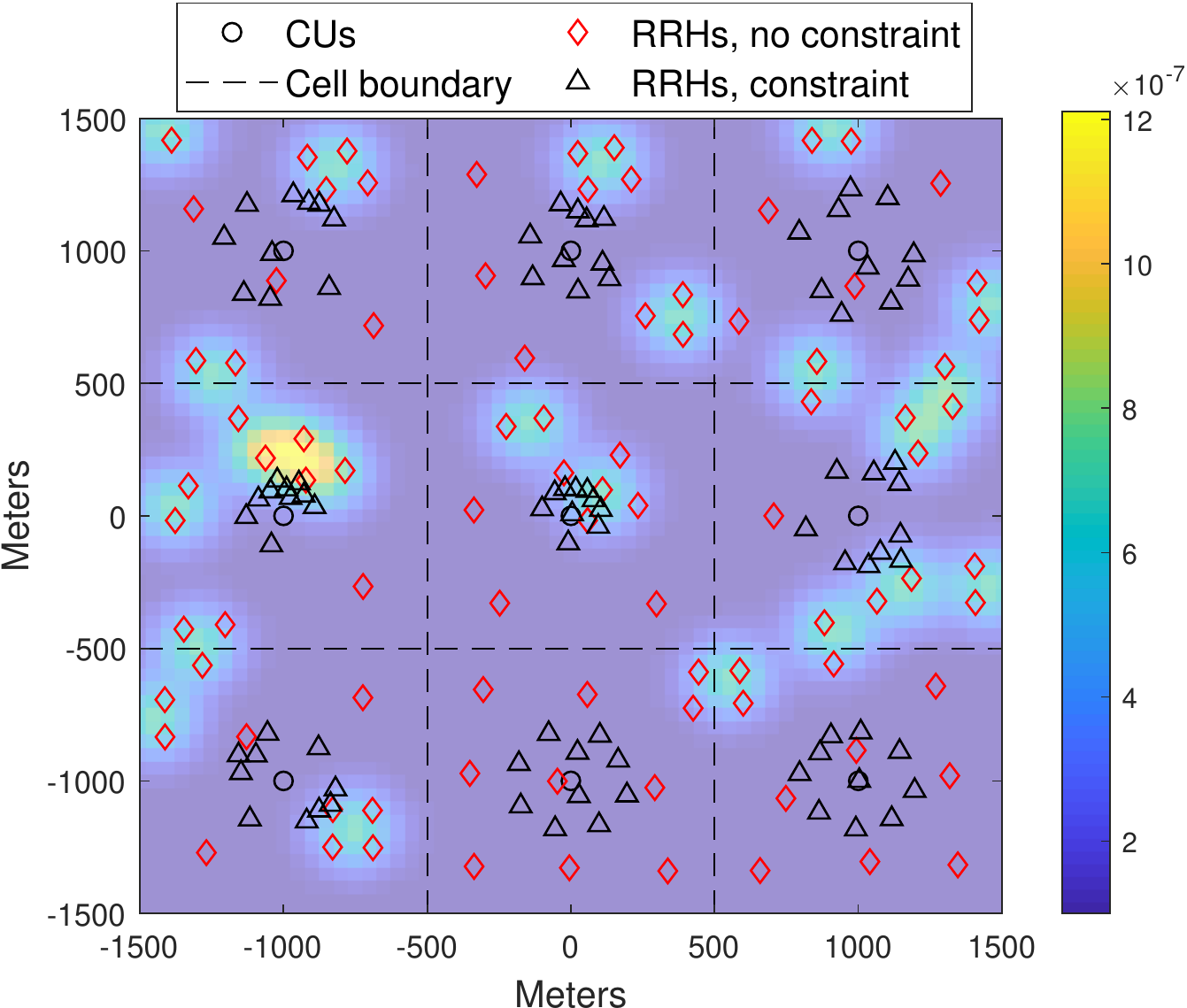}
	\caption{Typical generated network hotspots showing the optimal RRHs location when $\omega=6$ RBs, and at no constraint.}
	\label{fig:omega6noConstraintComparison}
	\vspace{-1em}
\end{figure}

In Figure \ref{fig:omega6noConstraintComparison}, we show a typical generated traffic distribution (equation~\eqref{eq:traffic_dist}), and we present the resulted optimized locations of the RRHs when we have $\omega = 6$ RBs (i.e., $\omega_c = 19$ RBs). We note that the users exist also in locations outside the hotspots (hotspots are represented with yellow areas) with a probability $P_0$ shown in Table~\ref{table:sim_parameters}. We include the no constraint case for comparison purpose. The results show that even at this high bandwidth allocation for the backhaul compared to the access channel, the backhaul constraint is still the limiting factor in the deployment of RRHs in each cell.

\begin{figure}[H]
	\centering
	\includegraphics[width=0.4\textwidth]{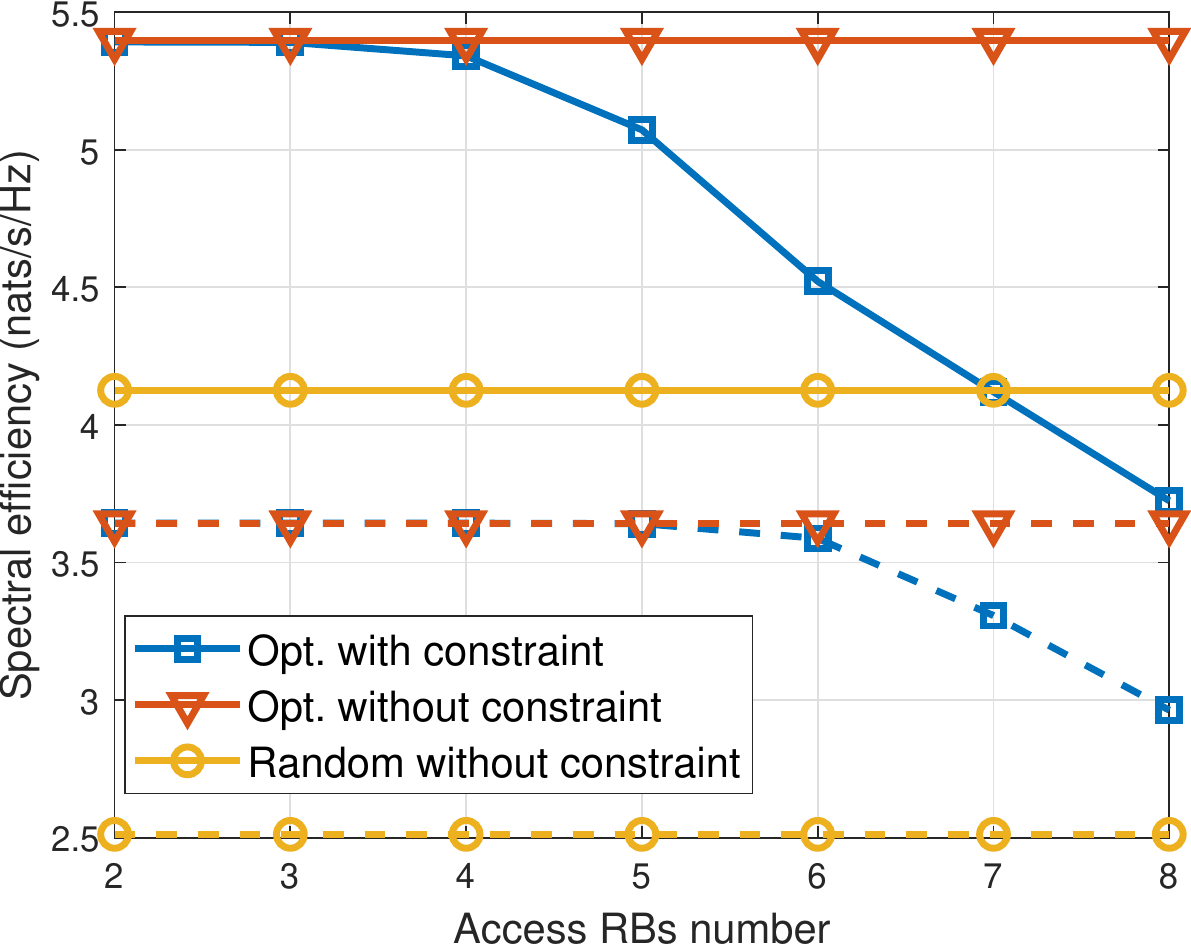}
	\caption{Achievable spectral efficiency as a function of different RBs allocation, $M = 8$ (solid line) and $M = 2$ (dashed).}
	\label{fig:BWallocation}
\end{figure}

In Figure \ref{fig:BWallocation}, we plot the spectral efficiency on the access channel as a function of different RBs allocations between the access channel and the backhaul. A ratio of $\frac{K\omega}{\omega_c}=1.36$ corresponding to $\omega \le 3$ RBs for the access channel allows deploying the RRHs freely in the network for the typical parameters specified in Table \ref{table:sim_parameters}. In such an allocation, the wireless backhaul is not a bottleneck and the RRHs can be freely deployed as if the backhaul has unlimited bandwidth. This constraint becomes even more relaxed if we use only $M=2$ antennas at each RRH, which leads to a lower spectral efficiency over the access channel and hence an $\omega \le 5$ RBs would be enough for freely deploying the RRHs.

\begin{figure}[H]
	\centering
	\includegraphics[width=0.4\textwidth]{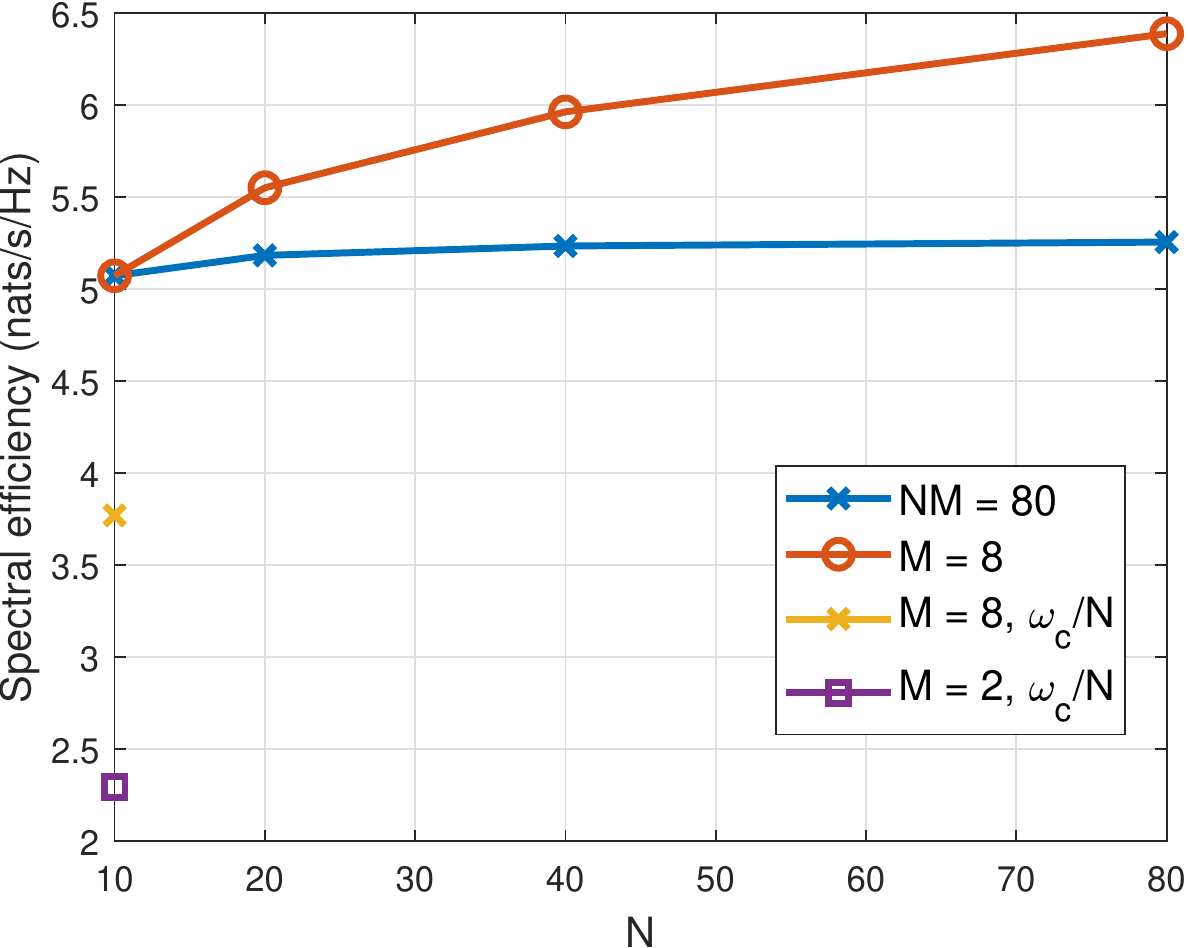}
	\caption{Access channel spectral efficiency at $\omega=5$ RBs.}
	\label{fig:plot_vs_N}
\end{figure}

In Figure \ref{fig:plot_vs_N}, we plot the spectral efficiency as a function of the number of the RRHs per cell, $N$. We show the results when the number of antennas $M$ per RRH is fixed ($M=8$), and when the total number of antennas per cell is fixed ($NM = 80$), i.e., as $N$ is increased we get more distributed network. Interestingly, the results show that distributing available antennas on more RRHs per cell is a good strategy to increase the spectral efficiency even when we have a backhaul constraint. On the other hand, we plot the obtained efficiencies when the backhaul bandwidth is further divided among the RRHs to provide frequency division among the RRHs links, i.e., the $w_c/N$ plots. For these plots, the backhaul constraint cannot be satisfied even when the RRHs are co-located at the CU, which means distributing the RRHs in the cell will not be possible, and other solutions should be taken to make such approach successful, e.g., decreasing $K$ or provide more bandwidth for the backhaul.
\vspace{-1em}
\vspace{-2em}
\vspace{-1.5em}

\section{Conclusion}\label{sec:Conclusion}
We analyzed the effect of a limited capacity backhaul on the achievable rate in a coordinated distributed network. We used point-to-point noise-limited wireless backhaul links between the CUs and the RRHs and analyzed its effect on the deployment decisions of the RRHs. We used ZF-BF on the access channel to allow the RRHs to jointly serve the users that are distributed according to some traffic distribution. Our results show that we need to allocate a very large bandwidth for the backhaul compared to the access channel to allow serving a large number of users and to allow free deployment of RRHs in the network. Our work underlines the fundamental role the backhaul plays in the design of distributed networks.
\appendix
\subsection{Proof of outage on the Noise-limited SISO Backhaul}
\label{appendix:RateConstraint}
The backhaul outage probability can be written as
\begin{align}\label{eq:P_outage_1}
&\displaystyle P_{qn}\left({\bf x},{\bf y}\right)=
\resizebox{0.38\textwidth}{!}
{$\displaystyle
\mathbb{P}\left\{R_{qn}^{(b)}\left(x_{qn},y_{qn}\right) \le
K \frac{\omega}{\omega_c} \mathbb{E}_{\widetilde{x}_k,\widetilde{y}_k}\left\{ R_{qk}^{(a)}({\bf x},{\bf y}) \right\}
\right\}
$}
\nonumber\\
&\ 
=
\resizebox{0.45\textwidth}{!}
{$\displaystyle
	\mathbb{P}\left\{\left|\bar{g}_{qn}\right|^2 \le \frac{1}{\rho_c \bar{\ell}_q(x_{qn},y_{qn})}\left(\exp\left(
	K \frac{\omega}{\omega_c}
	\mathbb{E}_{\widetilde{x}_k,\widetilde{y}_k}\left\{ R_{qk}^{(a)}({\bf x},{\bf y}) \right\}
	\right)-1\right)\right\}
	$}
\end{align}
where $\bar{g}_{qn}$ is the small-fading parameter for both the LoS and NLoS components between CU $q$ and RRH $n$, which is assumed to be Rician fading, hence the probability density function (PDF) of the fading power $\delta_{qn}=\left|\bar{g}_{qn}\right|^2$ is $
f(\delta_{qn})=\frac{1}{\eta_2^2}\exp\left(-\frac{\eta_1^2+\delta_{qn}}{\eta_2^2}\right)I_0\left(\frac{2\eta_1}{\eta_2^2}\sqrt{\delta_{qn}}\right)
$, where $I_0(z)=\frac{1}{\pi}\int_{0}^{\pi}\exp\left(z\cos(\theta)\right)\diff\theta$ is the modified Bessel function of first kind. 
Let us denote the right side of the inequality in~\eqref{eq:P_outage_1} as $\zeta_{qn}({\bf x},{\bf y})$, hence
\begin{align}
\displaystyle P_{qn}({\bf x},{\bf y})=
\resizebox{0.39\textwidth}{!}
{$\displaystyle
\int_{0}^{\zeta_{qn}({\bf x},{\bf y})}f(\delta_{qn})\diff \delta_{qn}=1-Q_1\left(\frac{\sqrt{2}\eta_1}{\eta_2},\frac{\sqrt{2 \zeta_{qn}({\bf x},{\bf y})}}{\eta_2}\right)
$}
\end{align}
where $Q_1(\cdot)$ is the Marcum Q-function. 

\section*{Acknowledgment}
This work was supported in part by Ericsson Canada and in part by the Natural Sciences and Engineering Research Council (NSERC) of Canada.

\ifCLASSOPTIONcaptionsoff
\newpage
\fi

\footnotesize
\bibliography{BD_conf_References}
\bibliographystyle{unsrt}

\end{document}